\documentstyle[12pt]{article}

\newcommand{\p}[1]{(\ref{#1})}
\begin{document}
\renewcommand{\thefootnote}{\fnsymbol{footnote}}
\thispagestyle{empty}
{\hfill  Preprint JINR E2-99-42}\vspace{0.5cm} \\

\begin{center}
{\large\bf Massless and spinning particles as dynamics in
one dimensional (super)diffeomorphism groups.}
 \vspace{1.5cm} \\
A. Pashnev\footnote{E-mail: pashnev@thsun1.jinr.dubna.su}\vspace{1cm}\\
{\it JINR--Bogoliubov Laboratory of Theoretical Physics} \\
{\it Dubna, Head Post Office, P.O.Box 79, 101 000 Moscow, Russia}
\vspace{1.5cm} \\
{\bf Abstract}
\end{center}
It is shown that dynamics of $D+2$ elements of the (super)diffeomorphism
group in one (1+1 for the super) dimension describes the $D$ - dimensional
(spinning) massless relativistic particles. The coordinates of this
elements ($D+2$ einbeins, $D+2$ connections and $1$ additional
common coordinate of higher dimensionality) play the role of
coordinates, momenta and Lagrange multiplier, needed for the
manifestly conformal and reparametrization invariant description
of the $D$ - dimensional (spinning) particle in terms
of the $D+2$ - dimensional spacetime.

\begin{center}
{\it Submitted to ``Classical and Quantum Gravity"}
\end{center}
\vfill
\setcounter{page}0
\renewcommand{\thefootnote}{\arabic{footnote}}
\setcounter{footnote}0
\newpage
\section{Introduction}
As is well known there exist several equivalent formulations of
the massless relativistic particles. The second order and first
order formalisms are examples of them. The essential ingredient of
both approaches is einbein - the field describing one dimensional
gravity. One more example is the conformally invariant description
\cite{M}, which starts from $D+2$ dimensional spacetime. The
existence of alternative approaches always sheds some new light on
the nature of the physical system. In particular, the conformally
invariant description from the very beginning considers the
particle coordinates and einbein on the equal footings. For the
extended spinning particle \cite{BCL}-\cite{PS} the analogous
description  \cite{S} shows that the gravitinos of the
corresponding one dimensional supergravity are on the same footing
with coordinates superpartners as well.

In the present work we consider the natural description of the
massless relativistic particle and $N=1$ spinning particle in terms of
nonlinear realization of the
infinite dimensional
diffeomorphisms group of the one dimensional
space ((1,1) superspace).

We construct the first order conformally invariant formulation
and show that the spacetime coordinates and one dimensional supergravity
fields are realized as dilatons of one dimensional diffeomorphisms
group. We consider simultaneously the dynamics of $D+2$
different points in the group space, hence they contain the same number
of dilatons. The corresponding $(D+2)$ components of momentum
are connected with the
Cristoffel symbols. One more parameter of the group having higher
dimension is the same for all $(D+2)$ points. It plays the role of
Lagrange multiplier and effectively
reduces the number of spacetime coordinates from $D+2$ to $D$ ones.

In the second section of the paper we shortly describe the conformally
invariant approach to relativistic particles and spinning particles.
The third section is devoted to the description of spinless
particle in terms of the diffeomorphisms group. 
In the fourth and fifth sections we construct the
reparametrization invariant in the $(1,1)$ superspace worldvolume
action for $N=1$ spinning particle. Some further possibilities
of applying the developed formalism are discussed in Conclusions.

\setcounter{equation}0\section{Conformally invariant description}
In this section for convenience of reader we remind the conformally
invariant description of the relativistic particle \cite{M}, $N=1$
\cite{M},\cite{S} and extended \cite{S} spinning particle.

The action for bosonic massless relativistic particle in
$D$ - dimensional
spacetime can be written in terms of $D+2$ coordinates
$x_{\cal A},\;{\cal A}=0,1,\ldots
D+1,$ of the spacetime with the signature
\begin{equation}\label{sigma}
\Sigma_{\cal A}=(-\underbrace{++\ldots ++}_{\mbox{$D$}}-):
\end{equation}
\begin{equation}               \label{1S}
S=\int d \tau(\frac{1}{2}\dot{x}^2-\frac{1}{2}\lambda x^2).
\end{equation}
Besides of the $SO(D,2)$ invariance, it is
gauge-invariant under the transformations
\begin{eqnarray}  \label{2S}
\delta x&=&\epsilon\dot{x} - \frac{1}{2}\dot{\epsilon}x,\\ \label{3S}
\delta\lambda&=&\epsilon\dot{\lambda}+2\dot{\epsilon}\lambda+
\frac{1}{2}\stackrel{\ldots}{\epsilon}.
\end{eqnarray}
The relation of the action \p{1S} with the usual $D$ - dimensional action
is established by solving the equation of motion for the Lagrange
multiplier $\lambda$
\begin{equation}
x^{\cal A}x_{\cal A}\equiv x^ax_a+2x_+x_-=0;\;\;\;a=0,\ldots,D-1;\;\;\;
x_\pm=\frac{1}{\sqrt{2}}(x_D\pm x_{D+1}).
\end{equation}
In terms of new variables
\begin{equation}\label{4S}
\tilde{x}=\frac{x}{x_+},\;\; \;e=\frac{1}{x_+^2}, \;\;\;
(x_-=-\frac{x^ax_a}{2x_+}),
\end{equation}
the Lagrangian becomes
\begin{equation}
L=\frac{1}{2}\;\frac{\dot{\tilde{x}}^2}{e}.
\end{equation}
Its reparametrization invariance
\begin{equation}
\delta \tilde{x}=\epsilon\dot{\tilde{x}},\;\;\;
\delta e=\dot{\epsilon}e+\epsilon\dot{e}
\end{equation}
is the consequence of \p{2S}-\p{3S}.

The modification of the action \p{1S} to the case of
extended spinning particle \cite{GT}, \cite{HPPT} is \cite{S}
\begin{equation}\label{ext}
L=\left(\frac{1}{2}{\dot x}^2+
\frac{1}{2}i{\dot\gamma}_i\cdot\gamma_i\right)-
\left(\frac{1}{2}\lambda x^2+i\lambda_i\gamma_i\cdot x+
\frac{1}{2}i\lambda_{ij}\gamma_i\cdot\gamma_j\right)
\end{equation}
(for $N=1$ spinning particle the action was constructed in \cite{M}).
Here $\gamma_i,\; i=1,\ldots N,$ are Grassmann variables which
become $\gamma$ - matrices upon quantization. After the solution
of the equations of motion for the Lagrange multipliers
$\lambda,\;\lambda_i$ and some redefinitions like \p{4S}
one can derive the usual $D$ - dimensional
action for $N$ - extended spinning particle in the form of 
\cite{GT}, \cite{HPPT}.

So, the Lagrange multipliers $\lambda$ in pure bosonic case and
$\lambda,\;\lambda_i$ in the case of extended spinning
particle play the crucial role in the conversion of the $D+2$ -
dimensional actions into $D$ - dimensional ones. Nevertheless, their
geometrical meaning as well as the nature of initial
$D+2$ coordinates $x_{\cal A}$ is unclear. 
In the next sections we will show
that all this functions of $\tau$ have an interpretation in terms
of parameters of diffeomorphisms groups.

\setcounter{equation}0\section{Geometrical description of the massless
particle}

Consider the auxiliary $1$ - dimensional bosonic space
with the coordinate $s$. The generators of the
corresponding diffeomorphisms group
\begin{equation}                \label{rep}
{L}_m=is^{m+1}\frac{\partial}{\partial s},\;
\end{equation}
form the Virasoro algebra without central charge
\begin{equation}\label{algebra}
\left[L_n,L_m\right]=-i(n-m)L_{n+m}.
\end{equation}
In what follows we will consider the subalgebra of the algebra \p{algebra}
which is formed by the regular at the origin
generators $L_m,\;\;m\geq -1.$

The most natural is the following parametrization of the group element
\begin{eqnarray}\label{coset}
&&G=e^{i\tau L_{-1}} \cdot e^{iU^{(1)}L_1} \cdot e^{iU^{(2)}L_2} \cdot
e^{iU^{(3)}L_3}\ldots e^{i{U}L_0},
\end{eqnarray}
in which all multipliers
with the exception of $e^{i{U}L_0}$, $\quad U\equiv U^{(0)}$,
are ordered by the dimensionality
of the correspondent generators: $\left[L_m\right]=m$.
Such structure of the group element simplifies the evaluation of the
variations $\delta {U^{(m)}}$ under the infinitesimal
left action
\begin{equation}\label{left}
G'=(1+i\epsilon)G,
\end{equation}
where $\epsilon =
\sum_{m=0}^\infty \epsilon^{(m)} L_{m-1}
$ belongs to the
algebra of the diffeomorphisms group.
The transformation laws of the coordinates in \p{coset} are \cite{P}
\begin{eqnarray}
\delta \tau &=&\varepsilon(\tau)\equiv\epsilon^{(0)}+\epsilon^{(1)}\tau+
\epsilon^{(2)}\tau^2+\ldots\;,\\
\delta U&=&\dot{\varepsilon}(\tau),\\
\delta {U^{(1)}}&=&
-\dot{\varepsilon}(\tau)U^{(1)}+
\frac{1}{2}\ddot{\varepsilon}(\tau),\\
\delta {U^{(2)}}&=&
-2\dot{\varepsilon}(\tau)U^{(2)}+
\frac{1}{6}\stackrel{\ldots}{\varepsilon}(\tau).
\end{eqnarray}
In general $U^{(n)}$ transforms through $\tau$ and $U^{(m)}, \; m<n$.
At this stage it is natural to consider all parameters as the fields
in one dimensional space parametrized by the coordinate $\tau$.
It means the following active form of the transformations of the
parameters $U(\tau), U^{(m)}(\tau)$
\begin{eqnarray}
\delta U(\tau)&=&-\varepsilon(\tau)\dot{U}(\tau)+
\dot{\varepsilon}(\tau),\\
\delta {U^{(1)}(\tau)}&=&-\varepsilon(\tau)\dot{U}^{(1)}(\tau)
-\dot{\varepsilon}(\tau)U^{(1)}(\tau)+
\frac{1}{2}\ddot{\varepsilon}(\tau),\\
\delta {U^{(2)}(\tau)}&=&-\varepsilon(\tau)\dot{U}^{(2)}(\tau)
-2\dot{\varepsilon}(\tau)U^{(2)}(\tau)+
\frac{1}{6}\stackrel{\ldots}{\varepsilon}(\tau).
\end{eqnarray}

One can easily verify that the functions
$x=e^{U(\tau)/2}$ and
$\lambda=-3U^{(2)}(\tau)$
have exactly the transformation laws \p{2S}-\p{3S}
with $\varepsilon(\tau)=-\epsilon$. Simultaneously $U^{(1)}(\tau)$
transforms as one dimensional Cristoffel symbol.

The independence of $\delta U^{(2)}$ from $U$ and $U^{(1)}$ means that one
can consider more than one group elements
\begin{equation}\label{ga}
G_{\cal A}=e^{i\tau L_{-1}}
 \cdot e^{iU^{(2)}L_2} \cdot
e^{iU^{(3)}L_3}\ldots \;\;
e^{iU_{\cal A}^{(1)}L_1} \cdot
e^{iU_{\cal A}L_0},\;\;{\cal A}=0,1\ldots,D+1,
\end{equation}
which have identical values of parameters
 $\tau$ and $U^{(m)}(\tau),\quad m\geq 2$, and
differ in the values of the parameters $U_{\cal A}^{(1)}$ and
$U_{\cal A}^{(0)}\equiv U_{\cal A}$.
This property is valid when all of these group elements are transformed with
the same infinitesimal transformation parameter $\varepsilon(\tau)$.

In general it is true for any group which admits the parametrization in
the form $G=K\cdot H$, where $H$ is some subgroup of the group $G$ and $K$
parametrizes   the
corresponding coset $K=G/H$. One can consistently consider the set
of group elements
\begin{equation}\label{general}
G_{\cal A}=K\cdot H_{\cal A}
 \end{equation}
  with equal
coset element $K$ and different elements of subgroup
$ H_{\cal A}$. This property (the equality of the coset elements
for all $G_{\cal A}$ ) is invariant with respect to the left
multiplication
\begin{equation}\label{generaltransform}
G_{\cal A} \rightarrow G'_{\cal A}=g\cdot G_{\cal A}
\end{equation}
with some group element $g$.

Consider the Cartan's differential form for each value of the index
$\cal A$
\begin{equation}
\Omega_{\cal A}=G_{\cal A}^{-1}dG_{\cal A}=i\Omega_{\cal A}^{(-1)}L_{-1}+
i\Omega_{\cal A}^{(0)}L_0+i\Omega_{\cal A}^{(1)}L_1+\ldots.
\end{equation}
All their components $(\Omega_{\cal A}^{(-1)},\;\Omega_{\cal A}^{(0)},\;
\Omega_{\cal A}^{(1)},\ldots)$
are invariant with respect to the left transformation 
\p{generaltransform}.
The explicit expressions for the components of the $\Omega$ -form are:
\begin{eqnarray}\label{a}
\Omega_{\cal A}^{(-1)}&=&e^{-U_{\cal A}}d\tau,\\  \label{ab}
\Omega_{\cal A}^{(0)}
&=&d U_{\cal A}-2d\tau U_{\cal A}^{(1)},\\ \label{abc}
\Omega_{\cal A}^{(1)}&=&(dU_{\cal A}^{(1)}+d\tau 
(U_{\cal A}^{(1)})^2-3d\tau
U^{(2)})e^{U_{\cal A}},\ldots\;.
\end{eqnarray}
The first of these forms is differential one-form einbein. The
covariant derivatives (carrying the external index 
${\cal A}$) calculated with its help are
\begin{equation}
D_{\tau\cal A}=e^{U_{\cal A}}\frac{d}{d\tau}.
\end{equation}

The most interesting is the form $\Omega_{\cal A}^{(1)}$.
The following expression for the action
\begin{eqnarray} \label{act}
S&=&-\frac{1}{2}\int\sum_{\cal A}\Sigma_{\cal A}\Omega_{\cal A}^{(1)}=\\
&=&-\frac{1}{2}\int d \tau\sum_{\cal A}\Sigma_{\cal A}e^{U_{\cal A}}
(\dot{U}_{\cal A}^{(1)}+ (U_{\cal A}^{(1)})^2-3U^{(2)}),\nonumber
\end{eqnarray}
where $\Sigma_{\cal A}$ is the signature \p{sigma}, is invariant
under the transformation \p{generaltransform} and
corresponds to the first order formalism for the action
\p{1S}. Indeed, after the integration by parts in the first term
and change of variables
\begin{equation}
x_{\cal A}=e^{U_{\cal A}(\tau)/2},\;\; p_{\cal A} =e^{U_{\cal A}(\tau)/2}
U_{\cal A}^{(1)},\;\;
\lambda=-3U^{(2)}(\tau)
\end{equation}
 it becomes (omitting summation over indices $\cal A$ with
 signature \p{sigma})
\begin{equation}\label{actfirst}
S_f=\int d \tau(\dot{x}p-\frac{1}{2}p^2-\frac{1}{2}\lambda x^2).
\end{equation}
This action is invariant under the gauge transformations
\begin{eqnarray}
\delta x&=&\epsilon\dot{x} - \frac{1}{2}\dot{\epsilon}x,\\
\delta\lambda&=&\epsilon\dot{\lambda}+2\dot{\epsilon}\lambda+
\frac{1}{2}\stackrel{\ldots}{\epsilon}.\\
\delta p&=&\epsilon \dot{p}+\frac{1}{2} \dot{\epsilon}p-
\frac{1}{2}\ddot{\epsilon}x.
\end{eqnarray}
After the elimination of $p_{\cal A}$
with the help of its equation of motion $p_{\cal A}=\dot{x}_{\cal A}$
the action \p{actfirst} coincides with the action \p{1S}.

\setcounter{equation}0
\section{$N=1$ spinning particle in a
superconformal gauge}
To generalize the approach on the spinning particles we firstly consider
more simple example of the
$N=1$ Superconformal Algebra (SCA)
\begin{eqnarray}
\left[L_m,L_n\right]&=&-i(m-n)L_{m+n}\\
\left[L_m,G_s\right]&=&-i(\frac{m}{2}-s)G_{m+s}\\
\{G_r,G_s\}&=&2L_{r+s}.
\end{eqnarray}
The indices $m,n\geq -1$ are integer and $r,s\geq -1/2$ -halfinteger.
Following the considerations of the previous Chapter and \cite{P}
we write the group element as
\begin{eqnarray}
G_{\cal A}&=&e^{i\tau L_{-1}} \cdot e^{i\theta G_{-1/2}}
 \cdot e^{i\Theta^{(3/2)}G_{3/2}}
 \cdot e^{iU^{(2)}L_2} \cdots \\
&&e^{i\Theta_{\cal A}G_{1/2}}  \cdot e^{iU_{\cal A}^{(1)}L_1}
 \cdot e^{iU_{\cal A}L_0},\;\;{\cal A}=0,1\ldots,D+1.\nonumber
\end{eqnarray}
Last three multipliers in this expression form the subgroup of the
whole superconformal group and they consistently can carry external
index ${\cal A}$, as discussed in the previous Chapter.

All parameters (Grassmann $\Theta$-s and commuting $U$-s)
are considered as superfunctions of $\tau$ and $\theta$
which parametrize the $(1,1)$ superspace. The variation of superspace
coordinates under the left action of infinitesimal
superconformal transformation can be written in terms of one
bosonic superfunction $\Lambda$
\begin{eqnarray}
\delta\tau&=& \Lambda-\frac{1}{2}\theta D_\theta\Lambda,\\
\delta\theta&=&-\frac{i}{2}D_\theta\Lambda,
\end{eqnarray}
where
\begin{equation}\label{flatD}
D_\theta=\frac{\partial}{\partial\theta}+
i\theta\frac{\partial}{\partial\tau},
\end{equation}
is the flat supercovariant derivative.

To calculate the invariant differential $\Omega$- forms one should
take into account that Grassmann parity of differential of any
variable is opposite to its own Grassmann parity, i.e.
$d\tau$ is odd and $d\theta$ is even \cite{B}.
The general expression for $\Omega$ -form is
\begin{equation}
\Omega_{\cal A}=G_{\cal A}^{-1}dG_{\cal A}=i\Omega_{\cal A}^{(-1)}L_{-1}+
i\Omega_{\cal A}^{(-1/2)}G_{-1/2}+i\Omega_{\cal A}^{(0)}L_0+
i\Omega_{\cal A}^{(1/2)}G_{1/2}+i\Omega_{\cal A}^{(1)}L_1+\ldots,
\end{equation}
where two first components
\begin{eqnarray}
\Omega_{\cal A}^\tau&\equiv&\Omega_{\cal A}^{(-1)}
=(d\tau-id\theta\theta)e^{-U_{\cal A}}=dx^M{E_M^\tau}_{\cal A} ,\\
\Omega_{\cal A}^\theta&\equiv&\Omega_{\cal A}^{(-1/2)}
=\{d\theta-(d\tau-id\theta\theta)\Theta_{\cal A}\}e^{-U_{\cal A}/2}=
dx^M{E_M^\theta}_{\cal A}
\end{eqnarray}
define supervielbein ($x^1\equiv\tau, x^2\equiv\theta):$\\
\begin{tabular}{c|ll|r}
  & $e^{-U_{\cal A}}$ & $-\Theta_{\cal A}\cdot e^{-U_{\cal A}/2}$&\\
$\phantom{aaaaaaaaaaaaaaaaaaaa}{E_M^A}_{\cal A}=$&&&\\
  & $-ie^{-U_{\cal A}}\cdot\theta$ &
$e^{-U_{\cal A}/2}(1-i\Theta_{\cal A}\cdot\theta)$&\\
\end{tabular}\vspace{0.3cm}\\
The covariant derivatives ${\cal D}_{B\cal A}
\equiv {E_B^M}_{\cal A}\partial_M,\quad
(B=\tau,\;\theta)$,
are defined
with the help of inverse supervielbein\\
\begin{tabular}{c|ll|}
  & $e^{U_{\cal A}}(1+i\Theta_{\cal A}\cdot\theta)$ &
 $\Theta_{\cal A}\cdot e^{U_{\cal A}}$\\
$\phantom{aaaaaaaaaaaaaaaaaaaa}{E_A^M}_{\cal A}=$&&\\
  & $ie^{U_{\cal A}/2}\cdot\theta$ &
$e^{U_{\cal A}/2}$\\
\end{tabular}\vspace{0.3cm}\\
As a result
\begin{equation}
{\cal D}_{\theta{\cal A}}=e^{U_{\cal A}/2}D_\theta,\;\;
{\cal D}_{\tau{\cal A}}=e^{U_{\cal A}}(D_\tau+\Theta_{\cal A}D_\theta),
\end{equation}
where $D_\tau\equiv\frac{\partial}{\partial\tau}$ and $D_\theta$
\p{flatD} are flat covariant derivatives.
The invariant integration measure is
\begin{equation}
dV_{\cal A}=d\tau\underline{d\theta}Ber({E_M^A}_{\cal A}),
\end{equation}
where $\underline{d\theta}$ is the Berezin differential and
\begin{equation}
Ber({E_M^A}_{\cal A})=e^{-U_{\cal A}/2}.
\end{equation}
Note, that all considered quantities -- supervielbein, covariant
derivatives and integration measure, depend on the external
 index ${\cal A}$.

To construct the action for $N=1$ spinning particle consider the
component $\Omega_{\cal A}^{(1)}$ and express it in terms of the full
system of invariant differential forms $\Omega_{\cal A}^\tau$ and
$\Omega_{\cal A}^\theta$
\begin{equation}
\Omega_{\cal A}^{1}=\Omega_{\cal A}^{\tau}Y_{\cal A}+
\Omega_{\cal A}^{\theta}\Gamma_{\cal A}.
\end{equation}
The coefficients are also invariant. In particular, $\Gamma_{\cal A}$
is odd and can be used for the construction of invariant action
\begin{eqnarray} \label{scact}
S&=&\frac{i}{2}\int \sum_{\cal A}dV_{\cal A}
\Sigma_{\cal A}\Gamma_{\cal A}=\\
&&\!\!\!\!\!\!\!\!\frac{i}{2}\int d\tau d\underline{\theta}
\sum_{\cal A}\Sigma_{\cal A}e^{U_{\cal A}}
(D_\theta U^{(1)}_{\cal A}-iD_\theta\Theta_{\cal A}
 \Theta_{\cal A}+
2i\Theta_{\cal A}U^{(1)}_{\cal A}-2i\Theta^{(3/2)}).
\end{eqnarray}
After the introduction of new variables
$$X_{\cal A}=e^{U_{\cal A}/2},\;\;\Pi_{\cal A}=
e^{U_{\cal A}/2}U^{(1)}_{\cal A},
\;\;\Xi_{\cal A}=e^{U_{\cal A}/2}\Theta_{\cal A},$$
integration by parts in first term and omitting the index ${\cal A}$
the action becomes
\begin{equation}
S=\frac{i}{2}\int d\tau d\underline{\theta}(-2\Pi D_\theta X-iD_\theta \Xi
\cdot \Xi+2i\Xi\Pi -2i\Theta^{3/2}\cdot  X^2).
\end{equation}
The equation of motion for $\Pi$ gives $\Xi=-iD_\theta X$.
Making use of this equation and identity $D_\theta^2=i\partial_\tau$
one can find the final result for the action in terms of even superfields
\begin{equation}
X_{\cal A}=x_{\cal A}+i\theta\gamma_{\cal A}
\end{equation}
and odd ones
\begin{equation}
\Theta^{3/2}=-\frac{1}{2}(\lambda_{odd}-\theta\lambda),
\end{equation}
\begin{equation}
S=-\frac{i}{2}\int d\tau d\underline{\theta}
(\dot{X}D_\theta X +2\Theta^{3/2} X^2)
\end{equation}
After the Berezin integration over $\theta$ it coincides with the
manifestly conformal component action for the $N=1$
spinning particle \p{ext}.

\setcounter{equation}0\section{Reparametrization invariant $N=1$
spinning particle}
The diffeomorphisms group of the superspace with one even and
one odd coordinates $s$ and $\eta$ is generated by two families
of even operators $N_n, \;n\geq -1$, and $M_m, \; m\geq 0$
\begin{equation}
N_n=is^{n+1}\frac{\partial}{\partial s},\;\;
M_n=is^n\eta\frac{\partial}{\partial \eta}
\end{equation}
 and
two families of odd operators $P_r,\;,Q_r,\;r\geq -1/2$
\begin{equation}
P_{n-1/2}=is^n\frac{\partial}{\partial \eta},\;\;
Q_{n-1/2}=is^n\eta\frac{\partial}{\partial s}.
\end{equation}.
Their
algebra is
\begin{eqnarray}
&&\left[N_m,N_n\right]=-i(m-n)N_{m+n},\\
&&\left[N_m,M_n\right]=inM_{m+n},\\
&&\left[N_m,P_s\right]=i(s+\frac{1}{2})P_{m+s},\\
&&\left[N_m,Q_s\right]=-i(m-s+\frac{1}{2})Q_{m+s},\\
&&\left[M_m,P_s\right]=-iP_{m+s},\\
&&\left[M_m,Q_s\right]=iQ_{m+s},\\
&&\{P_r,Q_s\}=iN_{r+s}+i(r+\frac{1}{2})M_{r+s}.
\end{eqnarray}
The superconformal algebra as subalgebra  is
generated by
\begin{equation}\label{sca}
  L_n=N_n+\frac{n+1}{2}M_n,\quad G_r=P_r-iQ_r.
\end{equation}
One can take the rest of linearly independent
generators in the form
\begin{equation}\label{rest}
 M_n,\quad F_r=P_r+iQ_r.
\end{equation}
It is convenient to write the group element as
\begin{eqnarray}
G_{\cal A}&=&e^{i\tau L_{-1}} \cdot e^{i\theta P_{-1/2}}
\cdot e^{i\psi Q_{-1/2}}\cdot
e^{iV^{(1)}M_1}
 \cdot
 e^{i\Theta^{3/2}P_{3/2}}
 \cdot e^{i\Psi^{3/2}Q_{3/2}}
 \cdots\\
&&
e^{i\Theta_{\cal A}P_{1/2}}  \cdot
e^{i\Psi_{\cal A}Q_{1/2}}  \cdot
e^{iU_{\cal A}^{(1)}L_1}  \cdot 
e^{iU_{\cal A}L_0}\cdot e^{iV_{\cal A}M_0}    ,
\;\;{\cal A}=0,1\ldots,D+1.\nonumber
\end{eqnarray}
The last five multipliers in this expression form the subgroup and
corresponding parameters $\Theta_{\cal A},\Psi_{\cal A},
U_{\cal A}^{(1)}, U_{\cal A}, V_{\cal A} $ carry additional external index
${\cal A}$.
All parameters (odd $\Theta, \Psi$ and even $U, V$)
again are considered as superfunctions of $\tau$ and $\theta$
which parametrize the $(1,1)$ superspace.
The left infinitesimal transformation leads to the reparametrization
of superspace coordinates
\begin{equation}
\delta\tau=a(\tau,\theta),\;\;\delta\theta=\xi(\tau,\theta)
\end{equation}
and to the following variation of $\psi$ \cite{P}
\begin{equation}
\delta\psi=-\partial_\theta a+\dot{a}\psi-\partial_\theta\xi\psi.
\end{equation}
One can show that such gauge freedom is enough to choose gauge
\begin{equation}\label{gauge}
\psi=-i\theta.
\end{equation}
Before going to such gauge one can calculate all invariant quantities -
supervielbein, covariant derivatives and integration measure:\\
\begin{tabular}{c|ll|r}
  & $e^{-U_{\cal A}}$ & $-\Theta_{\cal A}\cdot e^{-V_{\cal A}}$&\\
$\phantom{aaaaaaaaaaaaaaaaaaaa}{E_M}^A_{\cal A}=$&&&\\
  & $e^{-U_{\cal A}}\cdot\psi$ &
$e^{-V_{\cal A}}(1+\Theta_{\cal A}\cdot\psi)$&\\
\end{tabular}\vspace{0.3cm}\\
\begin{eqnarray}
{\cal D}_{\theta {\cal A} } &=&e^{V_{\cal A}}D_\theta,\;\;D_\theta=
\partial_\theta-\psi\partial_\tau,\\
{\cal D}_{\tau {\cal A} } &=&
e^{U_{\cal A}}(\partial_\tau+\Theta_{\cal A}
D_\theta),
\end{eqnarray}
\begin{equation}
Ber({E_M}^A_{\cal A})=e^{-U_{\cal A}+V_{\cal A}}.
\end{equation}
As in the case of superconformal algebra, described in the previous section,
consider the component of $\Omega$-forms corresponding to the generator
$L_1$
\begin{equation}
\Omega(L_1)_{\cal A}=\Omega_{\cal A}^{\tau}\cdot {\tilde Y}_{\cal A}+
\Omega_{\cal A}^{\theta}\cdot{\tilde  \Gamma}_{\cal A},
\end{equation}
and write the invariant action in the form
\begin{eqnarray}\label{difact}
S&=&\frac{i}{2}\int \sum_{\cal A}
d\tau \underline{d\theta}\cdot e^{-U_{\cal A}+V_{\cal A}}
\Sigma_{\cal A}{\tilde\Gamma}_{\cal A}=\\\nonumber
&&\frac{i}{2}\int d\tau \underline{d\theta}
\sum_{\cal A}\Sigma_{\cal A}\cdot e^{-U_{\cal A}+2V_{\cal A}}\{
D_\theta U^1_{\cal A}+D_\theta\Theta_{\cal A}\cdot
\Psi_{\cal A}+\Psi^{3/2}+D_\theta\psi\cdot \Theta^{3/2}-\\\nonumber
&& U^{(1)}_{\cal A}( \Psi_{\cal A}\cdot  +
D_\theta\psi\cdot \Theta_{\cal A})-V^{(1)}( \Psi_{\cal A}\cdot  -
D_\theta\psi\cdot \Theta_{\cal A})\}.
\end{eqnarray}

Obviously, the combination $\Psi^{3/2}+D_\theta\psi\cdot \Theta^{3/2}$
should be considered as one independent field.
The field $V^{(1)}$ in the action plays the role of Lagrange
multiplier which leads to the equation
\begin{equation}\label{cc}
\sum_{\cal A}\Sigma_{\cal A}\cdot 
 e^{-U_{\cal A}+2V_{\cal A}}\cdot(\Psi_{\cal A}  -
D_\theta\psi\cdot \Theta_{\cal A})=0.
\end{equation}
Note, that in the 
superconformal subgroup, generated by \p{sca}, 
takes plase more strong equation for 
each value of the index ${\cal A}$
$$(\Psi_{\cal A}  +i\Theta_{\cal A})=0$$
The equation \p{cc}, in contrast, contains the summation over 
the index ${\cal A}$.
Nevertheless, one can solve the constraint \p{cc} 
and substitute the solution
for  $\sum_{\cal A}\Sigma_{\cal A}\cdot 
e^{-U_{\cal A}+2V_{\cal A}}\cdot \Psi_{\cal A}$ back into 
the action \p{difact}.
In the resulting action without loss of any information one can 
choose the gauge
\p{gauge}. Indeed, when we calculate the equation of
motion for
$\psi$ and choose this gauge, they are consequence
of the equations of motion, which
follow from the gauge fixed action.

One easily can see that the gauge fixing reduces the action 
\p{difact} to the action
for $N=1$ superconformal group \p{scact}.
So, the action \p{difact}, which is invariant under the transformations
of the whole diffeomorphism group of the $(1,1)$ superspace
describes the $N=1$ spinning particle.

\setcounter{equation}0\section{Conclusions}

In the framework of nonlinear realizations of infinite - dimensional
diffeomorphism groups of one dimensional bosonic space and $(1,1)$
superspace we have constructed the conformally and reparametrization
invariant actions
for massless particle and $N=1$ spinning particle in arbitrary
dimension $D$. It is achieved by simultaneous consideration of
several group elements. The parameters of corresponding group points
include simultaneously the coordinates and momenta.
The interaction between coordinates is obliged to parameters
with higher dimensions, which are  the same for all considered
points on the group space.

It would be interesting to apply the method developed here and in \cite{P}
to other infinite dimensional symmetries, such as diffeomorphism groups
of extended superspaces and higher dimensional spaces, W-algebras and
so on

\noindent {\bf Acknowledgments.} I would like to thank Ch. Preitshopf
for useful discussions.

This investigation has been supported in part by the
Russian Foundation of Fundamental Research,
grant 99-02-18417,
joint grant RFFR-DFG 96-02-04022,
and INTAS, grants 93-127-ext, 96-0308,
96-0538, 94-2317 and grant of the
Dutch NWO organization.


\begin{thebibliography}{99}
\bibitem{M}R. Marnelius. Phys.Rev., {\bf D20} (1979) 2091
\bibitem{BCL}A. Barducci, R. Casalbuoni and L. Lusanna. Nuovo Cimento
            {\bf 35A} (1976) 377
\bibitem{GT}V.D. Gershun and V.I. Tkach. JETP Lett., 
            {\bf 29} (1979) 320
\bibitem{BDZVH}L. Brink, S Deser, B. Zumino, P. DiVecchia and P. Hove.
          Phys.Lett., {\bf 64B} (1976) 435
\bibitem{CT}P.A. Collins and R.W. Tucker. Nucl.Phys. {\bf B121} (1977) 307
\bibitem{BVH}L. Brink, P. DiVecchia and P.S. Howe. Nucl.Phys. {\bf B118}
           (1977) 76
\bibitem{HPPT}P. Howe, S. Penati, M. Pernici and P. Townsend.
         Phys.Lett., {\bf B215} (1988) 555
\bibitem{PS}A. Pashnev and D. Sorokin. Phys.Lett. {\bf B253} (1991) 301
\bibitem{S}W. Siegel. Int.J.Mod.Phys., {\bf A3} (1988) 2713
\bibitem{P}A. Pashnev.
Nonlinear realizations of the (super)diffeomorphism groups,
geometrical objects and integral invariants in the superspace.
Preprint JINR E2-97-122.\\ e-Print archive hep-th/9704203
\bibitem{B}F.A. Berezin. Introduction to the algebra and analysis with
anticommuting variables. Moscow Univ., 1983
\end{thebibliography}
\end{document}